\documentclass[conference]{IEEEtran}
\usepackage{cite}

\ifCLASSINFOpdf
 \usepackage[pdftex]{graphicx}
\else
\fi
\usepackage[cmex10]{amsmath}
\usepackage{algorithmic}
\usepackage{url}

\begin{document}
%
\title{Demographic and Structural Characteristics to Rationalize Link Formation in Online Social Networks}

\author{\IEEEauthorblockN{Muhammad Qasim Pasta}
\IEEEauthorblockA{Karachi Institute of Economics and Technology \\
Karachi, Pakistan\\
Email: mqpasta@pafkiet.edu.pk}


\and
\IEEEauthorblockN{Zohaib Jan}
\IEEEauthorblockA{Shaheed Zulfikar Ali Bhutto Institute of  Science and \\ Technology\\
Karachi, Pakistan\\
Email: zohaib.jan@szabist.edu.pk }


\and
\IEEEauthorblockN{Faraz Zaidi}
\IEEEauthorblockA{University of Lausanne, \\ 
Lausanne, Switzerland and \\
Karachi Institute of Economics and Technology \\
Karachi, Pakistan \\
Email: faraz@pafkiet.edu.pk}

\and
\IEEEauthorblockN{C\'{e}line Rozenblat}
\IEEEauthorblockA{University of Lausanne \\
Lausanne, Switzerland \\
Email: celine.rozenblat@unil.ch }
}

\maketitle

\begin{abstract}

Recent years have seen tremendous growth of many online social networks such as Facebook, LinkedIn and MySpace. People connect to each other through these networks forming large social communities providing researchers rich datasets to understand, model and predict social interactions and behaviors. New contacts in these networks can be formed either due to an individual's demographic profile such as age group, gender, geographic location or due to network's structural dynamics such as triadic closure and preferential attachment, or a combination of both demographic and structural characteristics.

A number of network generation models have been proposed in the last decade to explain the structure, evolution and processes taking place in different types of networks, and notably social networks. Network generation models studied in the literature primarily consider structural properties, and in some cases an individual's demographic profile in the formation of new social contacts. These models do not present a mechanism to combine both structural and demographic characteristics for the formation of new links. In this paper, we propose a new network generation algorithm which incorporates both these characteristics to model growth of a network. We use different publicly available Facebook datasets as benchmarks to demonstrate the correctness of the proposed network generation model.

\end{abstract}


%
\IEEEpeerreviewmaketitle

\section{Introduction}

Past decade has seen an exponential growth in the usage of online social networks such as Facebook, LinkedIn and MySpace \cite{boyd07} with hundreds of millions of users connecting to these networks everyday. The field of social network analysis and complex networks has profited from these networks as they provide rich datasets for researchers to investigate various hypothesis and conjectures related to social behavior and social dynamics in our society \cite{lievrouw02,garton97}. These networks in general, undergo several processes such as information propagation \cite{iribarren11}, marketing \cite{trusov08}, spreading viruses \cite{fan11} and community formation \cite{kumar06b} which can be studied using analysis methods, network metrics, visualization methods and clustering algorithms on large realistic datasets which was not possible in yesteryears.  

Substantial research has been conducted in modeling social networks where the objective has been to develop algorithmic models that can mimic structure and evolution of real world networks. More often than not, researchers have targeted structural characteristics such as high clustering coefficient, small geodesic distance, degree distribution following power-law, assortative mixing and presence of communities in these networks \cite{badham10,kumpula09,xu09,catanzaro04,holme02}.

These models are quite useful in the study of networks as they help to generate large networks with desired structural properties. Thus, giving us a better understanding of how networks are organized, how they evolve overtime and how structural dynamics impact the overall network properties. Furthermore, these models are also useful for simulation studies to examine different network processes taking place such as epidemic spread, influence mining and formation of community structures \cite{badham10,zaidi13b}. Another application area for these models is to test various sampling effects \cite{kurant11} as using these models, we can generate networks with different sizes and structural properties.

Apart from the structural characteristics, another aspect of these networks are the demographic characteristics of individuals that play an important role in the link formation. Demographic characteristics include attributes such as age group of an individual, gender, geographic location, professional activity sector, personal interests and hobbies \cite{preston01}. Most of the network generation models proposed in the literature do not consider these demographic characteristics. Some models have been proposed in the literature with the concept of social spaces and distances to refer to the demographic properties of individuals but the details of these properties are often omitted in these papers \cite{wong06,badham10}. They directly utilise distances drawn from some distribution to refer to how close two individuals are, which in turn determines the probability of link formation among individuals. We argue that it is to some extent, pivotal to consider both structural and demographic characteristics to develop a better understanding of the evolution process and rationalize link formation between two individuals in a network. 

In this paper, we propose a new network generation model, which considers both structural as well as demographic characteristics to generate social networks. The proposed algorithm is based on two steps: initialization and construction to generate networks with desired properties. We use different publicly available datasets from the famous social networking website Facebook to validate the proposed model as we were able to reproduce networks with similar properties. These results are documented in section \ref{sec:results}.

The rest of the paper is organized as follows: We discuss a number of articles that propose network generation models in section \ref{sec:related}. In section \ref{sec:proposedequation}, we formulate an equation to incorporate demographic as well as structural characteristics to determine similarity among two nodes, which in turn drives the connectivity of the whole network. In section \ref{sec:proposed}, we provide the details of the proposed model which consists of two steps, \textit{initialization} and \textit{construction}. Section \ref{sec:experimental} describes the experimental setup and the datasets used for comparative analysis followed by the results and explanation in section \ref{sec:results}. Finally, we conclude in section \ref{sec:conclusion} giving possible future research directions.

\section{Related Work}\label{sec:related}

The discovery of small world and scale free networks has revolutionized the way we study networks around us. Among other networks, social networks also exhibit small world and scale free properties. Watts and Strogatz (WS) \cite{watts98} proposed a model to simulate the occurrence of triadic closures (clustering coefficient) and the small world effect (short geodesic distances) in networks. Starting from a regular lattice, random rewiring of links with a certain probability \textit{p}, transforms a regular lattice into a network commonly known as small world networks. Albert and Barabasi (BA) \cite{barabasi99} introduced preferential attachment to simulate how networks with degree distribution following power-law evolve in real networks, commonly know as scale free networks or networks with scale free degree distribution. Starting from a few nodes, new nodes are introduced in the network which connect to older nodes with a probability proportional to the existing connectivity of the nodes. Nodes with higher degree have higher probability of forming new links, and these networks are commonly called scale free networks.

Most of the early works followed by these two ground breaking models revolved around the idea of having a unified model to generate both small world and scale free networks. For example, Holme and Kim \cite{holme02} proposed a modification to the BA model adding a triad formation step after the preferential attachment step to create triads in the network. This increases the overall clustering coefficient, thus generating a network with both small world and scale free properties. Other variants of the BA model such as \cite{dorogovtsev02,klemm02,guo05,fu06,wang08b,li12a} produce networks having high clustering coefficient by introducing triads one way or the other and nodes connect using the preferential attachment rule to have a scale free degree distribution.

Different researchers have used the idea of n-partite, and specially bi-partite graphs to generate social networks. The authors \cite{newman02a} introduce the idea to generate affiliation networks similar to co-authorship networks \cite{newman01} using random bipartite graphs with arbitrary degree distributions . This idea is also used by Guillaume and Latapy~\cite{guillaume04} as they identify bipartite graph structure as a fundamental model of complex networks by giving real world examples. The authors call  the two disjoint sets of a bipartite graph as \textit{bottom} and \textit{top}. At each step, a new \textit{top} node is added and its degree \textit{d} is sampled from a prescribed distribution. For each of the \textit{d} edges of the new vertex, either a new \textit{bottom} vertex is added or one is picked among the pre-existing ones using preferential attachment. The bipartite graph is then projected as a unipartite graph to obtain a small world and scale free network. A more generalized model based on similar principles was proposed \cite{bu07} where instead of using the bipartite structure, a network can contain \textit{t} disjoint sets (instead of just two sets, as is the case of the bipartite graph). The authors discuss the example of sexual web~\cite{lilijeros01} which is based on the bipartite structure. A sexual web is a network where nodes represent men and women having relationships to opposite sex, and similar nodes do not interact with each other. At each time step, a new node and \textit{m} new edges are added to the network with the sum of the probabilities equal to $1$. The preferential attachment rule is followed as the new node links with the existing nodes with a probability proportional to the degree of the nodes.

A growing network model \cite{catanzaro04} was proposed to incorporate the assortative mixing behavior in social networks. Assortative mixing here, refers to the structural property of individuals to connect with individuals having similar number of links. This model allows links to be added between existing individuals as well as new individuals on the basis of their degree thus forcing links between similar degree nodes, and inducing high assortativity in the network. 

Models based on demographic attributes have also been proposed where the goal is to determine connectivity based on social attributes. The social similarity, in these artefacts is often referred to as the social distance and the approach in general is termed as spatial approach for network generation. One such model based on social distance between individuals was presented by \cite{boguna04} where the model aims to generate networks with high clustering coefficient, assortativity and hierarchical community structures. Social distance refers to the degree of closeness or acceptance that an individual feels towards another individual in a social network. The closer two individuals are, the higher they have a probability to form a new link. The authors used a real acquaintance network to demonstrate the correctness of the proposed algorithm. Another model \cite{wong06} was proposed which uses spatial distance to model nodal properties and homophilic similarity among individuals. The model randomly spreads nodes in a geographical space such that the edge formation probability is dependent on the spatial distance among nodes. The network thus generated exhibits high clustering coefficient, small geodesic distance, power-law degree distribution, and the presence of community structures. 

A three phase spatial approach \cite{badham10} was proposed to generate networks with controllable structural parameters. This approach controls three important structural characteristics, the clustering coefficient, assortativity and degree distribution using input parameters making it quite useful to generate large networks. The model also takes as input, the degree sequence required in the final network. This static model uses a notational space to identify nodes closer to each other, a layout modification step to move nodes with similar degree closer and edge creation among nodes based on these spatial and layout modification step to achieve desired clustering coefficient and assortativity.

A very recent model focuses on the homophilic property of social networks \cite{almeida13}. The authors modify the BA model by introducing a homophilic term which creates regions where characteristics of individuals impact the rate of gaining links as well as links between individuals with similar and dissimilar characteristics. The model maintains five important network features, power-law degree distribution, preferential attachment, short geodesic distance, high clustering coefficient and growth over time.

Evolutionary network models with aging nodes have also been proposed in the literature such as \cite{dorogovtsev00a,zhu03,geng09,wen11}. For example \cite{wen11}, the authors study the dynamic behavior of weighted local-world evolving networks with aging nodes. Newly added nodes connect to existing nodes based on a strength-age preferential attachment and the results show that the network thus generated has power-law degree distribution, high clustering coefficient and small world properties.

There exists a number of models based on the local-world phenomena \cite{pan06,sun07,wang09,wen11} where nodes only consider there neighbourhood in contrast to traditional network models that assume the presence of global information. For example \cite{wang09} investigate a local preferential attachment model to generate hierarchical networks with tunable degree distribution, ranging from exponential to power-law.

Another class of graphs models, the exponential random graph models have gained a lot of popularity \cite{frank86,snijders06,robins07} also known as $\mathit{p^*}$. These models are used to test, to what extent nodal attributes and structural dependencies describe structure of a network measured using frequency of degree distribution, traids and geodesic distances \cite{toivonen09}. The possible ties among individuals are modelled as random variables, and assumptions about dependencies among these random tie variables determine the general form of the exponential random graph model for the network \cite{robins07}. An important difference between network generation models and ERGMs is that network models try to explain how a network evolves whereas ERGMs  do not explicitly explain network generation process \cite{toivonen09}.

Models to generate clustered graphs also exist in the literature where the goal is to have community structures embedded in the resulting networks \cite{condon99, lancichinetti09, moriano13, zaidi13a}. Since we do not address the issue of having community structures in the current work, we intend to incorporate this structural feature of many real world networks as part of future work.

An exhaustive review of network generation models is out of scope in this text, yet we have tried to cite a wide spectrum of different network generation models. Partial surveys, reports and comparative analysis for different network generation models can be found in  \cite{dorogovtsev02,newman03,fortunato09,badham10,toivonen09,zaidi13b}. None of the models to generate networks considers demographic and structural attributes during the network evolution process where as our contribution lies in considering demographic as well as structural characteristics as the driving force for link formation between individuals. The results we obtained from simulations using the proposed model demonstrate that the final networks obtained have small geodesic distances, high clustering coefficients and frequencies of degree distribution following power-law. We validate our model through comparative analysis as we generate networks similar to real world Facebook networks and the results are presented in section \ref{sec:results}. 



\section{Demographic and Structural Characteristics} \label{sec:proposedequation}

The proposed model is quite generic and aims to provide a general equation which can be further refined by adding more network specific details. First we introduce the equation, and then we provide details of the model implemented using the equation.

The premise upon which the proposed equation is developed is that, for individuals $i$ and $j$, the link formation is a function $\mathit{f}$ of two types of characteristics, demographic ($\mathit{D}$) and structural ($\mathit{S}$). Mathematically we can represent this relation as:

\begin{equation}
\mathit{f}(i,j)= \alpha \{ \mathit{D}_{i,j} \} + \beta \{ \mathit{S}_{i,j} \} 
\label{eq:f}
\end{equation}

where $\mathit{D}_{i,j}$ and $\mathit{S}_{i,j}$ represent the demographic and structural similarities between individual $i$ and $j$ respectively, $\alpha$ and $\beta$ represent equilibrium factors to control the balance between demographic and structural characteristics. Within this basic framework, different demographic and structural attributes can be considered. Specially for demographic characteristics, we propose a method to handle categorical, ordinal and numerical attributes separately, which can further be modified and tweaked depending upon the type of network to be generated, the available attribute information and other domain level knowledge that can be incorporated to justify link formation among pair of individuals. We discuss details of how demographic and structural characteristics are handled to calculate the possibility of link formation between two nodes below:

\subsection{Demographic Characteristics}\label{sec:demographic}

As discussed above, we consider different categorical, ordinal and numerical characteristics as demographic characteristics of an individual. For every categorical attribute $\mathit{C_p}$ where $p$ represents different attributes, the similarity between individuals $i$ and $j$ is assigned using the following equation:

\begin{equation}
\mathit{C_p}(i,j)=
\begin{cases}
1, \text{if $i_p=j_p$ }
\\
0, \text{if $i_p \neq j_p$ }
\end{cases}
\end{equation}

Similarly for every ordinal attribute $\mathit{O_q}$ where $q$ represents different attributes, the similarity between $i$ and $j$ is calculated using:

\begin{equation}
\mathit{O_q}(i,j)= \dfrac{|i_q - j_q|}{\mathit{\rho_q}} 
\end{equation}

where $i_q,j_q$ are the ranking orders, $|*|$ represents absolute value and $\mathit{O_q}$ is normalized using the maximum different ordinal values possible for attribute $q$ denoted by $\mathit{\rho_q}$ in the above equation. Similar to ordinal attributes, we calculate the normalized difference between numerical attributes of $i$ and $j$ using the following equation: 

\begin{equation}
\mathit{N_r}(i,j)= \dfrac{|i_r - j_r|}{\mathit{\rho_r}} 
\end{equation}

Using the above equations, we can calculate an accumulative similarity value using equations 1,2 and 3, based on weighted demographic characteristics as follows where $\omega$ represents weights associated to each attribute signifying its importance in the process of link formation.

\begin{equation}
\mathit{D}_{i,j}= \omega_p \mathit{C_p} + \omega_q \mathit{O_q} + \omega_r \mathit{N_r}
\label{eq:d}
\end{equation}

The above equation shows a linear combination of a categorical, an ordinal and a numerical characteristic to give a general form where any number of such demographic attributes can be combined together.

\subsection{Structural Characteristics}\label{sec:structural}

In case of structural characteristics, we consider two properties, the triadic closures (commonly known as friend-of-a-friend phenomena in sociology) which controls the global clustering coefficient, and preferential attachment to control the degree distribution of the generated network. Preference for formation of triadic closures as $i$ and $j$ have common friends is calculated using the following equation:

\begin{equation}
\mathit{FoF}(i,j)=  \dfrac{ i \  {\cap} \ j }  { min (i,j) } 
\end{equation}

where $i \ {\cap} \ j$ represents the common friends of i and j and $min(i,j)$ represents the minimum number of friends of either i or j. The minimum value in the denominator ensures that a relationship is not penalized just because one of the individual has high number of links. The more friends two individuals have in common, the more chances they have of forming a new link among themselves. As the network continuously evolves and new edges are added among previously added individuals, this process results in increasing overall clustering coefficient. To handle the preferential attachment $\mathit{PA}$ in link formation, we use the following equation: 

\begin{equation}
\mathit{PA}(i,j)= \dfrac{deg_i}{max(deg_n)} 
\end{equation}

For an newly added node $j$ (which initially will have zero connections), the probability of connecting to a node $i$ already existing in the network is directly proportional to the normalized degree of node $i$. The degree is normalized using the maximum node degree in the current network represented by $max(deg_n)$. We normalize this factor just to control the weight of each structural characteristic as all our characteristics are normalized between values 0 and 1. 


\begin{equation}
\mathit{S}_{i,j}= \omega_{FoF} \mathit{FoF} + \omega_{PA} \mathit{PA} 
\label{eq:s}
\end{equation}

Finally combining equation \ref{eq:d} and \ref{eq:s} as input to equation \ref{eq:f}, we can calculate an accumulated similarity for link formation between two individuals where both demographic as well as structural attributes are taken into account. Collectively, we refer to demographic and structural attributes as similarity based link formation.

\section{Proposed Model} \label{sec:proposed}

Apart from the distribution of demographic attributes, the model takes as input, the desired number of nodes in the network $n$, the minimum and maximum node degree $m_o$ and $ m_f$, the probability of similarity based link formation using equation \ref{eq:f} $P(Sim)$, the probability of triad formation $P(T)$ and triad count $t_c$ to determine the number of links that would be used to form triads. We also take weights $\omega$ for each demographic and structural attribute which can eventually help us to tune each characteristic's role in the formation of links among individuals. 

\begin{figure}
\begin{center}
\includegraphics[width=0.49\textwidth]{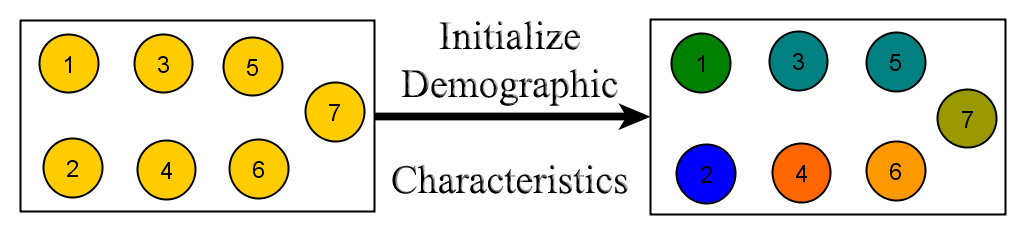}
\end{center}
\caption{The initialization step where nodes are randomly assigned demographic characteristics. Nodes are colored according to a combination of different characteristic values where similar colors represent similarity of nodes in terms of demographic characteristics.}
\label{fig::initialization}
\end{figure}

\begin{figure*}
\begin{center}
\includegraphics[width=0.79\textwidth]{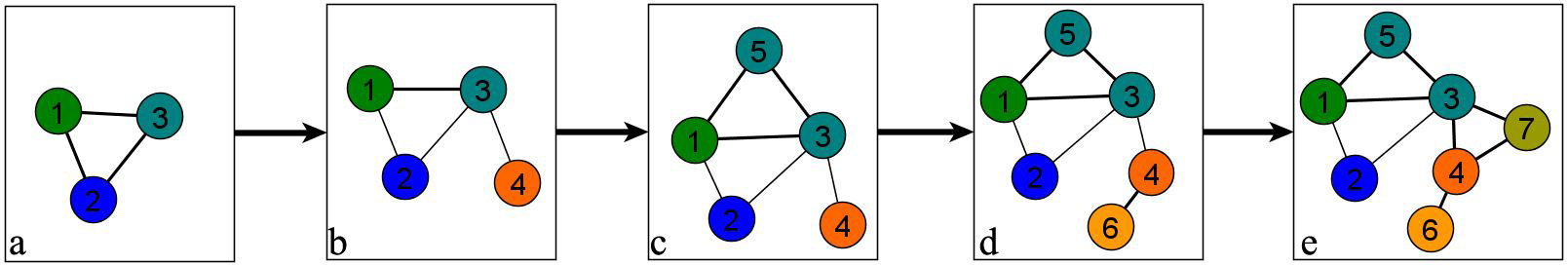}
\end{center}
\caption{Construction steps from (a) to (e) where initialized nodes are linked together based on demographic and structural characteristics. Every iteration adds a new node from initialized step and determines its similarity to existing nodes to possibly form links, and then possibly performs a triad formation step to create triads.}
\label{fig::construction}
\end{figure*}

The model comprises of two basic steps, the initialization step and the construction step. Within the construction step, two steps are performed, similarity based linking and triad formation. All these steps are described below:


\begin{enumerate}
\item \label{initialize} The initialization step randomly assigns demographic attributes in the given proportion to each of the $n$ nodes of the network. This results in a set of initialized nodes as shown in figure \ref{fig::initialization}. The nodes are numbered to associate a logical order which can be assigned randomly as the model is independent of this ordering of nodes.

\item \label{step1construction} To start construction of the connected network, the algorithm selects the first three nodes and connects them as a triad, irrespective of their similarity, as shown in figure \ref{fig::construction}(a).

\item \label{step2construction} A new node $n$ is then selected from the set of initialized nodes. A random number $m$ is generated between $m_o$ and $m_f$ to determine the number of edges of node $n$. While the total links of $n$ are less than $m$, the following two steps are repeated: 

\begin{enumerate}

\item \label{similaritybsed}Based on the probability of similarity based link formation $P(Sim)$, it connects to similar nodes in the construction phase based on similarity calculated through equation \ref{eq:f}. For example if the probability of the similarity based connection is 0.6, then the rest of the times $n$ is connected to a randomly selected node.

\item \label{triadformation}Based on the probability of triad formation $P(T)$, $n$ is then linked to $t_c$ neighbors of the nodes it connected to in the previous step, selecting the most similar nodes using equation \ref{eq:f} forming triads. For example if the probability of triad formation is 1, and $t_c$ is 2 then $n$ connects to two neighbors of the node it connected to in the previous step. In case, there are no neighbors, it chooses nodes randomly.

\end{enumerate}

\item \label{repeat} The process is repeated from step \ref{step2construction} until all the nodes in the initialization set are processed in the construction step.

\end{enumerate}

For clarification, we consider a small example with seven nodes. We consider the case of three demographic attributes, school (categorical), major (categorical) and age (numerical). Given as input, there are 3 possible schools in the proportion (2:2:3), there are two possible majors in the proportion (3:4) and the students have 3 possible age values in the proportion (3:3:1).
These attributes are assigned randomly to all the seven nodes as shown in figure \ref{fig::initialization} where the color coding in the initialization set depicts a unique color for a combination of attributes. So for nodes 3 and 5, the same color means that these individuals have exactly the same values for all demographic characteristics.

During the construction step, nodes from the initialization step are iteratively added to the network as shown in figure \ref{fig::construction}. Step (a) in figure \ref{fig::construction} shows that nodes 1, 2 and 3 are connected as a triad. Step (b) shows that node 4 is added to the network and connects to node 3 based on node similarity. Subsequently nodes 5,6 and 7 are added to the network where similar nodes form links on the basis of equation \ref{eq:f} and triad formation step introduces traids in the network.

\section{Data sets and Experimental Setup} \label{sec:experimental}

We used Facebook datasets provided by \cite{traud11} which represent the structure of 100 different american colleges and universities at a single point in time. The demographic attributes present in the dataset are gender, class year, major and residence (housing). We used five randomly choosen networks out of these hundred datasets for comparative analysis. The five networks are named as Caltech (769 nodes), Reed (962 nodes), Simmons (1518 nodes), Middlebury (3075 nodes) and American (6386 nodes) networks. 

We tested our model to simulate networks of exactly the same size as that of these five networks and the distribution of demographic attributes was kept exactly equal to the original datasets. As a result, the nodes have exactly the same distribution of demographic attributes as in the original networks. We perform a structural comparison the original and the generated networks using density, geodesic distances, clustering coefficient, power-law fit and assortativity. The power-law fit is calculated used the method proposed by \cite{clauset09}. The five graphs were generated using the parameters listed in table \ref{tbl:parameters}.

\begin{table}
\centering
\begin{tabular}{|l|c|c|c|c|c|}
\hline 
Dataset     & Min    & Max    & Probability of          & Probability of   & Triad \\
            & Edges  & Edges  & Links using             & Triad            & Count \\
            &        &        & Similarity              & Formation        &       \\
            & $m_o$  & $m_f$  & $P(Sim)$                & $P(T)$           & $t_c$ \\
\hline 
Caltech	    & 1      & 44     & 1                 & 1              & 3     \\
\hline
Reed	    & 1      & 40     & 1                 & 1              & 3     \\
\hline
Simmons	    & 1      & 43     & 1                 & 1              & 4     \\
\hline
Middlebury  & 1      & 83     & 1                 & 1              & 4     \\
\hline
American    & 1      & 72     & 1                 & 1              & 4     \\
\hline
\end{tabular}
\vspace{5pt}
\caption{Parameters used to generate graphs equivalent to original datasets from Facebook.} 
\label{tbl:parameters}
\end{table}

For the current experiments, the $\omega$ for all attributes is kept $1$, giving equal importance to all attributes. We plan to conduct an extensive study of the effects of varying these parameters and generating graphs with varying structural properties as part of future work.

\section{Results and Discussion}\label{sec:results}

\begin{figure}
\begin{center}
\includegraphics[width=0.3\textwidth]{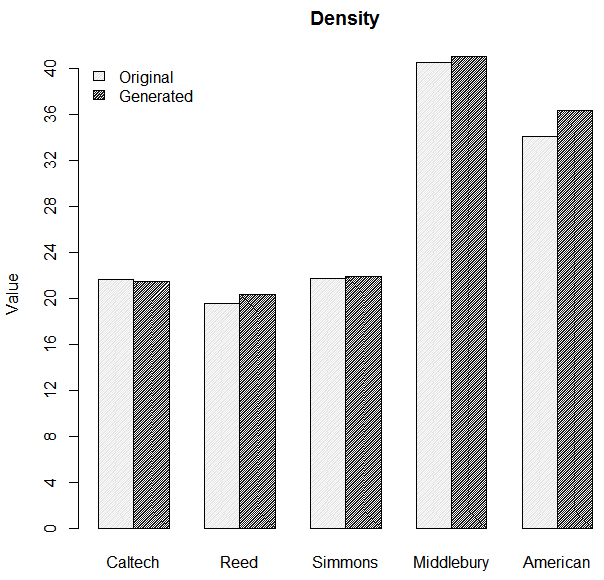}
\end{center}
\vspace{-10pt}
\caption{Comparative analysis of node-edge ratio or density of the original graphs and the generated graphs.}
\label{fig::density}
\end{figure}

\begin{figure}
\begin{center}
\includegraphics[width=0.3\textwidth]{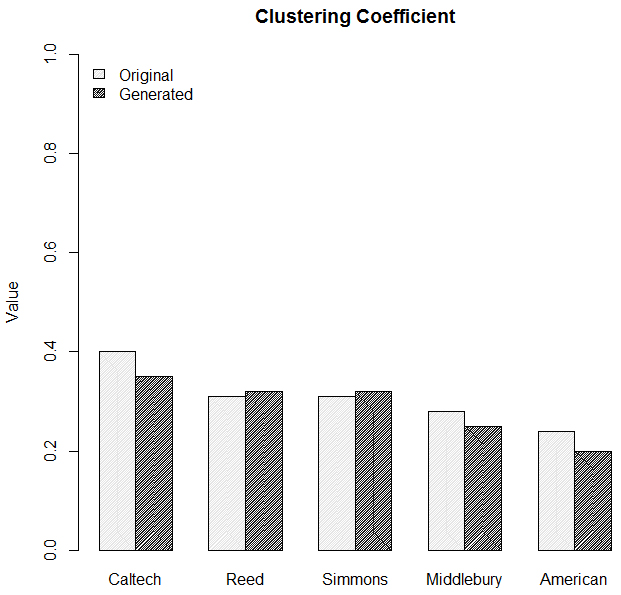}
\end{center}
\vspace{-10pt}
\caption{Comparative analysis of clustering coefficient of the original graphs and the generated graphs.}
\label{fig::cc}
\end{figure}

\begin{figure}
\begin{center}
\includegraphics[width=0.3\textwidth]{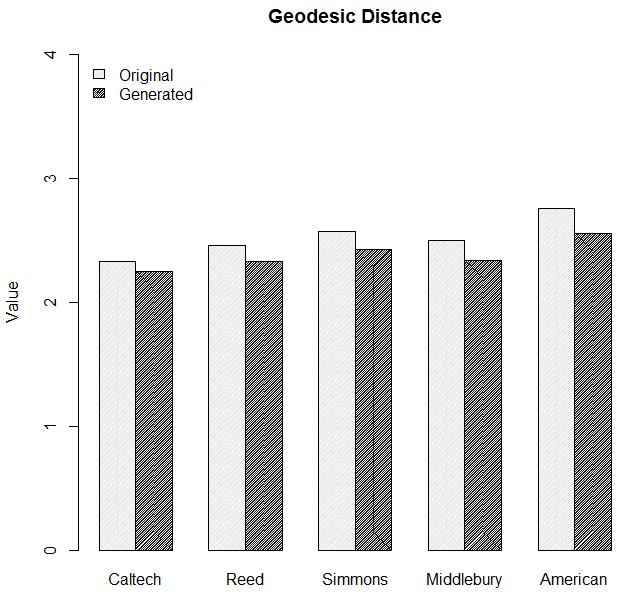}
\end{center}
\vspace{-10pt}
\caption{Comparative analysis of geodesic distances of the original graphs and the generated graphs.}
\label{fig::apl}
\end{figure}

\begin{figure}
\begin{center}
\includegraphics[width=0.3\textwidth]{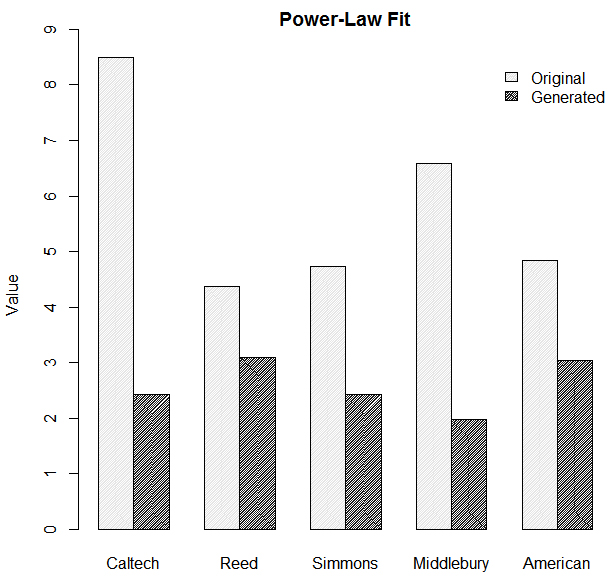}
\end{center}
\vspace{-10pt}
\caption{Comparative analysis of the power-law fit of the original graphs and the generated graphs.}
\label{fig::alpha}
\end{figure}

\begin{figure}
\begin{center}
\includegraphics[width=0.3\textwidth]{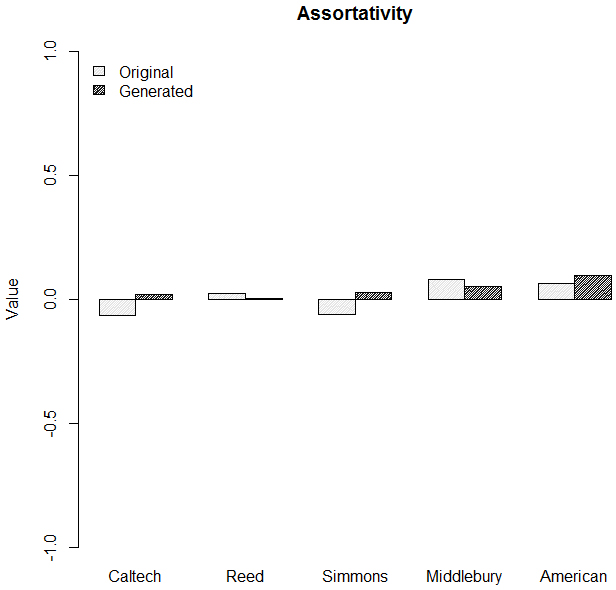}
\end{center}
\vspace{-10pt}
\caption{Comparative analysis of assortativity of the original graphs and the generated graphs.}
\label{fig::assor}
\end{figure}

We compared the generated graphs with the original graphs using five metrics, the node-edge ratio often called density, the clustering coefficient, the average geodesic distance, the power-law fit and assortativity. The results are shown in figure \ref{fig::density}, \ref{fig::cc}, \ref{fig::apl}, \ref{fig::alpha} and \ref{fig::assor} where the five datasets are compared to the generated networks using the proposed model.

In case of density, the values generated by the proposed model are very similar to the original networks as shown in figure \ref{fig::density}. The proposed model uses the parameters $m_o$ and $m_f$ where the mean of the two approximately represents the overall density of the generated network. Increasing these values increase the overall density and vice versa. An important remark about these parameters is that this does not necessarily mean that the maximum degree of a node will not exceed $m_f$. These parameters signify the number of connections that a new entering node will form, not with whom they form so it is normal that due to preferential attachment, a new node might connect to a node with very high degree which might have connections more than $m_f$.

Figure \ref{fig::cc} shows the clustering coefficients of the original and the generated graphs. Again, we were able to generate values that are very close to the desired values. The clustering coefficient is controlled through the parameters $P(T)$ and $t_c$ where $P(T)$ is the probability of triad formation taking place and $t_c$ represents the number of such triads to be formed. Increasing this number increases the overall clustering coefficient of the generated network.

In figure \ref{fig::apl}, we compare the geodesic distances of the networks again showing high similarity. We do not have any specific parameter to control this value but while calculating similarity based link formation, we consider preferential attachment based on degree connectivity, which results in both small geodesic distances for the generated graphs and their degree distribution following power-law as shown in figure \ref{fig::alpha}. All the generated networks have a power-law fit between $1.9$ and $3.1$ suggesting scale free behavior of the proposed model. We were not able to match the power-law fit with that of the original facebook networks, since we incorporated the preferential attachment model \cite{barabasi99}, which is known to result in scale free degree distributions with power-law fit around $2$ or $3$. This fact is also well known for social networks but with the facebook datasets we used, the values of power-law fit are not between $2$ or $3$. Our experimentation suggests that we need to modify the existing methods to generate degree distributions to have a better fit rather than using the known preferential attachment model. One way to achieve a matching degree distribution is to use the model proposed by \cite{molloy95} which generates a network given a degree distribution.

Figure \ref{fig::assor} shows the comparative assortativity values for the original and the generated networks. In case of Caltech and Simmons datasets, the original networks show a slightly negative assortativity, or disassortative mixing, where as generated networks although have also very small values, but they are still positive. In case of Reed, Middlebury and American datasets, the original as well as generated networks have all positive values. The differences between original and generated networks for all five datasets are negligible. The proposed model does not currently enforces any structural method to control assortativity in the generated networks but still the model was able to achieve very similar values to that of real Facebook datasets. 

\begin{figure}
\begin{center}
\includegraphics[width=0.3\textwidth]{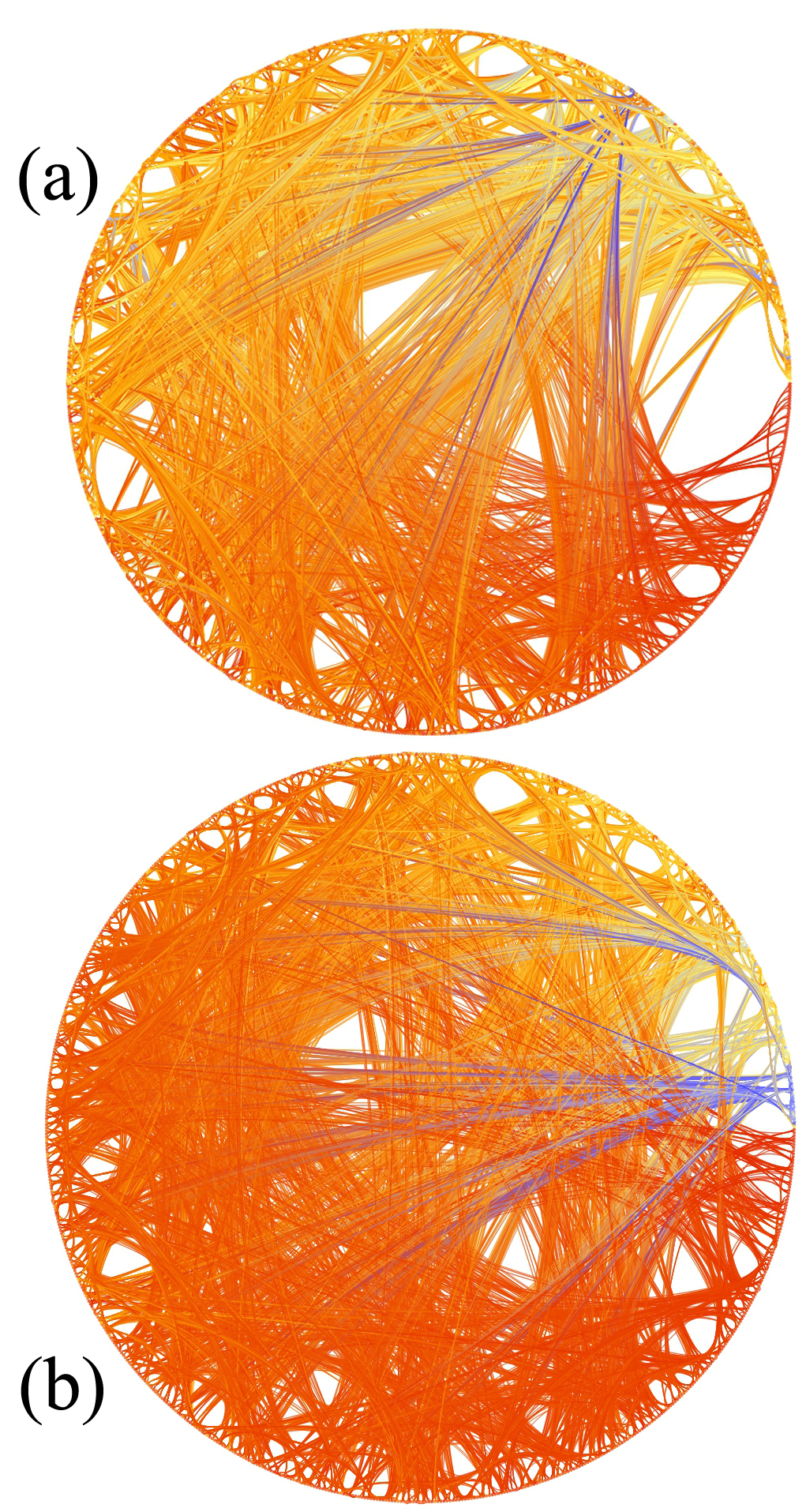}
\end{center}
\caption{Visual layout of the original and generated networks for the Caltech data set. The algorithms are layed out using circular layout. Nodes are colored with respect to node degrees with a gradient from Blue (High degree nodes) to Orange (Low degree nodes). The images are generated using Tulip Software \cite{auber03a}.}
\label{fig::caltechcomp}
\end{figure}

We also performed a visual comparison of the Caltech dataset which is the smallest network among the five networks with 769 nodes. Figure \ref{fig::caltechcomp} shows the layout of the the original network and the generated network using the proposed model. The nodes in the figure are colored with their degree. The nodes are placed on approximately the same locations in a circular layout. It is difficult to draw concrete conclusions about the similarity of each node but there are no major differences in the overall structure of the two networks. Since we did not intend to produce exactly the same network, we do not perform a node level comparison between the two networks. Furthermore, the proposed network model can be used to generate large size networks with similar structural and demographic properties, in which case, node level comparison will become meaningless.

Our comparative analysis shows high structural similarity among the original and generated networks apart from the power-law fit. The models is flexible any of the structural or demographic characteristic can be ignored (by assigning $\omega=0$), or given more importance (by reducing $\omega$ of other attributes). This flexibility is an important feature of the proposed model as it gives more control on how the network is generated as well as enables us to study the effects of different structural and demographic attributes.

To demonstrate the scalability of the proposed model, we generated different large size networks on a standard Intel i5 machine, 2.5 GHz dual core processor with 4GB memory. The running time in seconds for the generated networks are shown in Table \ref{tbl:running}.

\begin{table}
\centering
\begin{tabular}{|c|c|}
\hline 

Dataset Size     & Running Time \\
  Nodes          & Seconds      \\      
\hline 
1000            &  1            \\
10000           & 99            \\
100000          & 10305         \\
\hline
\end{tabular}
\vspace{5pt}
\caption{Running times for generating large size networks using the proposed model.} 
\label{tbl:running}
\end{table}

\section{Conclusion}\label{sec:conclusion}

In this paper, we have proposed a network generation model based on demographic and structural characteristics in order to better understand and rationalize link formation among individuals. We used different Facebook datasets to validate our model as it was successfully able to regenerate the same densities, clustering coefficients and geodesic distances. The model generated networks that are scale free using preferential attachment, but was unable to produce the same power-law fit as the original networks.

Extensive study needs to be performed to study the effects of $\omega$ which were kept $1$ through out our experiment as well as the balancing factors $\alpha$ and $\beta$. We intend to continue this study as part of future work to have a more generalized view of the proposed model. Furthermore, we have not included any structural characteristics to control assortative mixing of individuals and we plan to incorporate this feature as well, which will enable the current model to generate assortative as well as disassortative networks. Another important feature of social and other complex networks is the presences of community structures and we also foresee this amendment to the proposed model to generate more realistic networks.

\bibliographystyle{IEEEtran}
\bibliography{visu}

\begin{thebibliography}{10}
\providecommand{\url}[1]{#1}
\csname url@samestyle\endcsname
\providecommand{\newblock}{\relax}
\providecommand{\bibinfo}[2]{#2}
\providecommand{\BIBentrySTDinterwordspacing}{\spaceskip=0pt\relax}
\providecommand{\BIBentryALTinterwordstretchfactor}{4}
\providecommand{\BIBentryALTinterwordspacing}{\spaceskip=\fontdimen2\font plus
\BIBentryALTinterwordstretchfactor\fontdimen3\font minus
  \fontdimen4\font\relax}
\providecommand{\BIBforeignlanguage}[2]{{%
\expandafter\ifx\csname l@#1\endcsname\relax
\typeout{** WARNING: IEEEtran.bst: No hyphenation pattern has been}%
\typeout{** loaded for the language `#1'. Using the pattern for}%
\typeout{** the default language instead.}%
\else
\language=\csname l@#1\endcsname
\fi
#2}}
\providecommand{\BIBdecl}{\relax}
\BIBdecl

\bibitem{boyd07}
D.~Boyd and N.~B. Ellison, ``Social network sites: Definition, history, and
  scholarship,'' \emph{Journal of Computer-Mediated Communication}, vol.~13,
  no. 1-2, Nov. 2007.

\bibitem{lievrouw02}
L.~A. Lievrouw and S.~Livingstone, \emph{Handbook of new media: Social shaping
  and consequences of ICTs}.\hskip 1em plus 0.5em minus 0.4em\relax Sage, 2002.

\bibitem{garton97}
L.~Garton, C.~Haythornthwaite, and B.~Wellman, ``Studying online social
  networks,'' \emph{Journal of Computer-Mediated Communication}, vol.~3, no.~1,
  pp. 0--0, 1997.

\bibitem{iribarren11}
J.~L. Iribarren and E.~Moro, ``Affinity paths and information diffusion in
  social networks,'' \emph{Social Networks}, vol. 33 (2), pp. 134--142, 2011.

\bibitem{trusov08}
M.~Trusov, R.~E. Bucklin, and K.~H. Pauwels, ``Effects of {Word-of-Mouth}
  versus traditional marketing: Findings from an internet social networking
  site,'' \emph{Social Science Research Network Working Paper Series}, May
  2008.

\bibitem{fan11}
W.~Fan and K.~Yeung, ``Online social networks—paradise of computer viruses,''
  \emph{Physica A: Statistical Mechanics and its Applications}, vol. 390,
  no.~2, pp. 189 -- 197, 2011.

\bibitem{kumar06b}
R.~Kumar, J.~Novak, and A.~Tomkins, ``Structure and evolution of online social
  networks,'' in \emph{KDD '06: Proceedings of the 12th ACM SIGKDD
  international conference on Knowledge discovery and data mining}.\hskip 1em
  plus 0.5em minus 0.4em\relax New York, NY, USA: ACM, 2006, pp. 611--617.

\bibitem{badham10}
J.~Badham and R.~Stocker, ``A spatial approach to network generation for three
  properties: degree distribution, clustering coefficient and degree
  assortativity,'' \emph{Journal of Artificial Societies and Social
  Simulation}, vol.~13, no.~1, p.~11, 2010.

\bibitem{kumpula09}
J.~Kumpula, J.-P. Onnela, J.~Saram{\"a}ki, J.~Kertesz, and K.~Kaski, ``Model of
  community emergence in weighted social networks,'' \emph{Computer Physics
  Communications}, vol. 180, no.~4, pp. 517--522, 2009.

\bibitem{xu09}
X.-J. Xu, X.~Zhang, and J.~Mendes, ``Growing community networks with local
  events,'' \emph{Physica A: Statistical Mechanics and its Applications}, vol.
  388, no.~7, pp. 1273--1278, 2009.

\bibitem{catanzaro04}
M.~Catanzaro, G.~Caldarelli, and L.~Pietronero, ``Assortative model for social
  networks,'' \emph{Physical Review E (Statistical, Nonlinear, and Soft Matter
  Physics)}, vol.~70, no.~3, pp. 1--4, 2004.

\bibitem{holme02}
P.~Holme and B.~J. Kim, ``Growing scale-free networks with tunable
  clustering,'' \emph{Physical Review E}, vol.~65, p. 026107, 2002.

\bibitem{zaidi13b}
A.~Sallaberry, F.~Zaidi, and G.~Melan\c{c}on, ``Model for generating artificial
  social networks having community structures with small-world and scale-free
  properties,'' \emph{Social Network Analysis and Mining}, vol.~3, pp.
  597--609, 2013.

\bibitem{kurant11}
M.~Kurant, M.~Gjoka, C.~T. Butts, and A.~Markopoulou, ``Walking on a graph with
  a magnifying glass: stratified sampling via weighted random walks,'' in
  \emph{Proceedings of the ACM SIGMETRICS joint international conference on
  Measurement and modeling of computer systems}.\hskip 1em plus 0.5em minus
  0.4em\relax ACM, 2011, pp. 281--292.

\bibitem{preston01}
S.~H. Preston, P.~Heuveline, and M.~Guillot, ``Demography: Measuring and
  modeling population processes,'' \emph{Pop. Dev. Rev}, vol.~27, p. 365, 2001.

\bibitem{wong06}
L.~H. Wong, P.~Pattison, and G.~Robins, ``A spatial model for social
  networks,'' \emph{Physica A: Statistical Mechanics and its Applications},
  vol. 360, no.~1, pp. 99--120, 2006.

\bibitem{watts98}
D.~J. Watts and S.~H. Strogatz, ``Collective dynamics of 'small-world'
  networks,'' \emph{Nature}, vol. 393, pp. 440--442, Jun. 1998.

\bibitem{barabasi99}
A.~L. Barab\'{a}si and R.~Albert, ``Emergence of scaling in random networks,''
  \emph{Science}, vol. 286, no. 5439, pp. 509--512, 1999.

\bibitem{dorogovtsev02}
S.~N. {Dorogovtsev} and J.~F.~F. {Mendes}, ``Evolution of networks,''
  \emph{Advances in Physics}, vol.~51, pp. 1079--1187, Jun. 2002.

\bibitem{klemm02}
K.~Klemm and V.~M. Eguiluz, ``Growing scale-free networks with small world
  behavior,'' \emph{Physical Review E}, vol.~65, p. 057102, 2002.

\bibitem{guo05}
J.-G. Liu, Y.-Z. Dang, and Z.~tuo Wang, ``Multistage random growing small-world
  networks with power-law degree distribution,'' \emph{Chinese Phys. Lett.},
  vol.~23, no.~3, p. 746, Oct.~31 2005.

\bibitem{fu06}
P.~Fu and K.~Liao, ``An evolving scale-free network with large clustering
  coefficient,'' in \emph{ICARCV}.\hskip 1em plus 0.5em minus 0.4em\relax IEEE,
  2006, pp. 1--4.

\bibitem{wang08b}
J.~Wang and L.~Rong, ``Evolving small-world networks based on the modified ba
  model,'' \emph{Computer Science and Information Technology, International
  Conference on}, vol.~0, pp. 143--146, 2008.

\bibitem{li12a}
Y.~Li, X.~Qian, and D.~Wang, ``Extended hk evolving network model,'' in
  \emph{Control and Decision Conference (CCDC), 2012 24th Chinese}.\hskip 1em
  plus 0.5em minus 0.4em\relax IEEE, 2012, pp. 4095--4097.

\bibitem{newman02a}
M.~E.~J. Newman, D.~J. Watts, and S.~H. Strogatz, ``Random graph models of
  social networks,'' \emph{Proceedings of the National Academy of Sciences of
  the United States of America}, vol.~99, no. Suppl 1, pp. 2566--2572, February
  2002.

\bibitem{newman01}
M.~E. Newman, ``Scientific collaboration networks. i. network construction and
  fundamental results.'' \emph{Phys Rev E Stat Nonlin Soft Matter Phys},
  vol.~64, no. 1 Pt 2, July 2001.

\bibitem{guillaume04}
J.-L. Guillaume and M.~Latapy, ``Bipartite graphs as models of complex
  networks,'' in \emph{Workshop on Combinatorial and Algorithmic Aspects of
  Networking (CAAN), LNCS}, vol.~1, 2004.

\bibitem{bu07}
S.~Bu, B.-H. Wang, and T.~Zhou, ``Gaining scale-free and high clustering
  complex networks,'' \emph{Physica A: Statistical Mechanics and its
  Applications}, vol. 374, pp. 864--868, 2007.

\bibitem{lilijeros01}
F.~Lilijeros, C.~Edling, L.~Amaral, E.~Stanley, and Y.~{\aa}berg, ``The web of
  human sexual contacts,'' \emph{Nature}, vol. 411, pp. 907--908, 2001.

\bibitem{boguna04}
M.~Bogu{\~n}{\'a}, R.~Pastor-Satorras, A.~D\'{i}az-Guilera, and A.~Arenas,
  ``Models of social networks based on social distance attachment,''
  \emph{Physical Review E}, vol.~70, no.~5, p. 056122, 2004.

\bibitem{almeida13}
M.~L. de~Almeida, G.~A. Mendes, G.~M. Viswanathan, and L.~R. da~Silva,
  ``Scale-free homophilic network,'' \emph{The European Physical Journal B},
  vol.~86, no.~2, pp. 1--6, 2013.

\bibitem{dorogovtsev00a}
S.~N. Dorogovtsev and J.~F.~F. Mendes, ``Evolution of networks with aging of
  sites,'' \emph{Physical Review E}, vol.~62, no.~2, pp. 1842--1845, 2000.

\bibitem{zhu03}
H.~Zhu, X.~Wang, and J.-Y. Zhu, ``Effect of aging on network structure,''
  \emph{Physical Review E}, vol.~68, no.~5, p. 056121, 2003.

\bibitem{geng09}
X.~Geng and Y.~Wang, ``Degree correlations in citation networks model with
  aging,'' \emph{Europhysics Letters}, vol.~88, no.~3, p. 38002, 2009.

\bibitem{wen11}
G.~Wen, Z.~Duan, G.~Chen, and X.~Geng, ``A weighted local-world evolving
  network model with aging nodes,'' \emph{Physica A: Statistical Mechanics and
  its Applications}, vol. 390, no.~21, pp. 4012--4026, 2011.

\bibitem{pan06}
Z.~Pan, X.~Li, and X.~Wang, ``Generalized local-world models for weighted
  networks,'' \emph{Physical Review E}, vol.~73, no.~5, p. 056109, 2006.

\bibitem{sun07}
X.~Sun, E.~Feng, and J.~Li, ``From unweighted to weighted networks with local
  information,'' \emph{Physica A: Statistical Mechanics and its Applications},
  vol. 385, no.~1, pp. 370--378, 2007.

\bibitem{wang09}
L.-N. Wang, J.-L. Guo, H.-X. Yang, and T.~Zhou, ``Local preferential attachment
  model for hierarchical networks,'' \emph{Physica A: Statistical Mechanics and
  its Applications}, vol. 388, no.~8, pp. 1713--1720, 2009.

\bibitem{frank86}
O.~Frank and D.~Strauss, ``Markov graphs,'' \emph{Journal of the american
  Statistical association}, vol.~81, no. 395, pp. 832--842, 1986.

\bibitem{snijders06}
T.~A. Snijders, P.~E. Pattison, G.~L. Robins, and M.~S. Handcock, ``New
  specifications for exponential random graphs models,'' \emph{Sociological
  Methodology}, vol.~36, no.~1, pp. 99--153, Dec. 2006.

\bibitem{robins07}
G.~Robins, P.~Pattison, Y.~Kalish, and D.~Lusher, ``An introduction to
  exponential random graph (p) models for social networks,'' \emph{Social
  Networks}, vol.~29, no.~2, pp. 173--191, May 2007.

\bibitem{toivonen09}
R.~Toivonen, L.~Kovanen, M.~Kivelä, J.-P. Onnela, J.~Saramäki, and K.~Kaski,
  ``A comparative study of social network models: Network evolution models and
  nodal attribute models,'' \emph{Social Networks}, vol.~31, no.~4, pp. 240 --
  254, 2009.

\bibitem{condon99}
A.~Condon and R.~M. Karp, ``Algorithms for graph partitioning on the planted
  partition model,'' \emph{Random Structures and Algorithms}, vol. 18(2), pp.
  116--140, 1999.

\bibitem{lancichinetti09}
A.~Lancichinetti and S.~Fortunato, ``Benchmarks for testing community detection
  algorithms on directed and weighted graphs with overlapping communities,''
  \emph{Physical Review E}, vol.~80, no.~1, p. 016118, 2009.

\bibitem{moriano13}
P.~Moriano and J.~Finke, ``On the formation of structure in growing networks,''
  \emph{arXiv preprint arXiv:1301.4192}, 2013.

\bibitem{zaidi13a}
F.~Zaidi, ``Small world networks and clustered small world networks with random
  connectivity,'' \emph{Social Network Analysis and Mining}, vol. Volume 3,
  no.~1, pp. 51--63, 2013.

\bibitem{newman03}
M.~E.~J. Newman, ``The structure and function of complex networks,'' \emph{SIAM
  Review}, vol.~45, p. 167, 2003.

\bibitem{fortunato09}
\BIBentryALTinterwordspacing
S.~Fortunato, ``Community detection in graphs,'' Jun 2009. [Online]. Available:
  \url{http://arxiv.org/abs/0906.0612}
\BIBentrySTDinterwordspacing

\bibitem{traud11}
A.~L. Traud, P.~J. Mucha, and M.~A. Porter, ``Social structure of facebook
  networks,'' \emph{Physica A: Statistical Mechanics and its Applications},
  vol. 391, no.~16, pp. 4165--4180, 2011.

\bibitem{clauset09}
A.~Clauset, C.~R. Shalizi, and M.~E. Newman, ``Power-law distributions in
  empirical data,'' \emph{SIAM review}, vol.~51, no.~4, pp. 661--703, 2009.

\bibitem{molloy95}
M.~Molloy and B.~Reed, ``A critical point for random graphs with a given degree
  sequence,'' \emph{Random Structures and Algorithms}, vol.~6, pp. 161--180,
  1995.

\bibitem{auber03a}
D.~Auber, ``Tulip - a huge graph visualization framework,'' in \emph{Graph
  Drawing Software}, ser. Mathematics and Visualization Series, P.~Mutzel and
  M.~J\"{u}nger, Eds.\hskip 1em plus 0.5em minus 0.4em\relax Springer Verlag,
  2003.

\end{thebibliography}

\end{document}